\documentclass[aps,prx,twocolumn,english,showpacs,letterpaper,superscriptaddress]{revtex4-1}

\usepackage[latin9]{inputenc}
\usepackage{amssymb}
\usepackage{graphicx}
\usepackage{subfigure}
\usepackage{array}

\usepackage{epsfig}
\usepackage[hidelinks]{hyperref}

\usepackage{amsmath}
\usepackage{color}
\usepackage{float}
\usepackage{mathrsfs}
\usepackage{indentfirst}
\usepackage{textcomp}
\usepackage{comment}
\usepackage{mathtools}

\newcommand{\be}{\begin{equation}}
\newcommand{\ee}{\end{equation}}
\newcommand{\bea}{\begin{eqnarray}}
\newcommand{\eea}{\end{eqnarray}}

\newcommand{\p}{\partial}
\newcommand{\s}{\sigma}

\newcommand{\la}{\langle}
\newcommand{\ra}{\rangle}
\newcommand{\rd}{\mbox{d}}
\newcommand{\ri}{\mbox{i}}
\newcommand{\re}{\mbox{e}}

\begin{document}
\title{Simulating  exotic phases of matter with bond-directed interactions with arrays of Majorana-Cooper pair boxes}
\author{A.M. Tsvelik}
\affiliation{  Condensed Matter Physics and Materials Science Division, Brookhaven National Laboratory,
  Upton, NY 11973-5000, USA} \date{\today } 
\begin{abstract} 
It is suggested that networks of Majorana-Cooper pair boxes connected by metallic nanowires can simulate various exotic states of matter. In this simulations Majorana-Cooper boxes play the role of effective spins S=1/2 and the metallic connections generate the Kondo screening and the Ruderman-Kittel-Kasuya-Yosida (RKKY) interaction. Depending on what prevails - whether it is the Kondo effect  or the RKKY exchange, one will have either an effective spin model or a Kondo lattice.  The list  of exotic stets includes  the famous hexagonal Kitaev model, a generalization of this model for a Kondo lattice and various spin models with three-spin interactions. A special emphasize is made on the discussion of the Kondo lattice scenario. 
 \end{abstract}


\maketitle
%
%

 Search for phases of matter beyond one dimension which would support fractionalized excitations meets with  practical difficulties. As a rule existence of such phases requires peculiar interactions such as bond-directed or multi-spin exchange. Even  if  such interactions are present in realistic systems they may not be dominant ones as required by the theory.  One well known example of that kind is the bond-directed exchange in the celebrated Kitaev  model \cite{kitaev} - the model of spins S=1/2 on hexagonal lattice interacting with the bond-directed exchange  interaction:  
 \be
 H_{Kitaev} = J \sum_{{\bf r},{\bf e}}S^{\mu}({\bf r})S^{\mu}({\bf r}+ {\bf e}_{\mu}), \label{kitaev}
 \ee
 where ${\bf e}_{\mu}$ are the three vectors directed along the bonds of honeycomb lattice. As was demonstrated in \cite{kitaev}, this model describes  an algebraic spin liquid, its  propagating excitations are collective modes of spins - gapless Majorana fermions. As was pointed out in  \cite{jackeli} such bond-directed exchange does exist in real materials, for example, in RuCl$_3$ and Na$_2$IrO$_3$\cite{Chun2015}. It turns out, however, that it exists alongside with  other interactions including the ordinary Heisenberg exchange which mask the manifestation of  the pure spin liquid physics \cite{exp,exp2,review1}. 
 
 Another example of an exotic interaction cited in the literature  is the three-spin one 
 \be
 H_{three} = \frac{\chi}{2}\sum_{i,j,k \in \Delta}\Big({\bf S}_i[{\bf S}_j\times{\bf S}_k]\Big) \label{RKKY}
 \ee 
It has been suggested \cite{baskaran, zee} that such interaction would stabilize the Chiral Spin Liquid (CSL) state first described by Kalmeyer and Laughlin \cite{kl,kl2}. CSL  is an analogue of the $\nu =1$ Quantum Hall state in spin systems. This idea has been further developed in \cite{ludwig,ludwig2} and more recently in \cite{ferraz}. This spin singlet state breaks both time-reversal and parity symmetry; it shares the basic properties of quantum Hall states, such as a bulk spectral gap and chiral edge states \cite{kl,kl2,fradkin,yang}. Experimental realizations of such state are yet to be found. The main difficulty here is to find realistic situations where the three-spin interaction would be dominant.

\begin{figure}[!htb]
\centerline{\includegraphics[ angle = 0,
width=0.3\columnwidth]{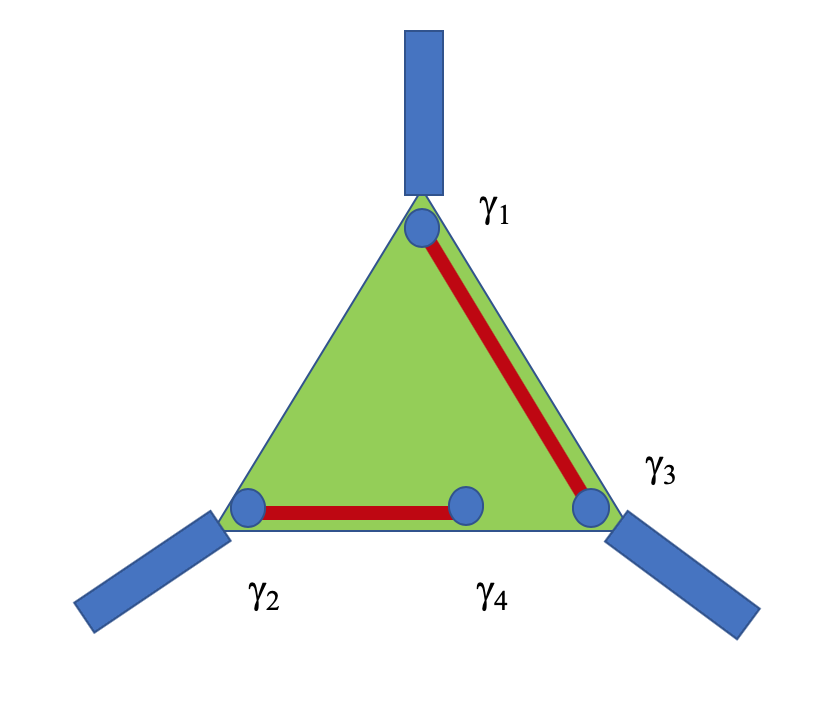}}
\vspace{-.50cm}
\caption{A schematic depiction of MCB with four Majorana zero energy modes (blue dots). Three modes are coupled to external metallic contacts (in blue), one mode remains idle. The green triangle is the superconducting (Cooper pair) box   with charging energy $E_C$. 
}
 \label{MCB}
\end{figure}

 I suggest that  {\it similar}  interactions and possibly many other exotic ones  can be generated in systems where effective spins S=1/2 are made artificially using Majorana-Cooper pair boxes (MCB).  Each MCB contains two nanowires made with a semiconductor with a strong spin-orbit interaction proximitized to a mesoscopic superconductor with charging energy $E_C$ (see Fig. \ref{MCB}). It was suggested in  \cite{lutchin, oreg} that with a suitable choice of parameters (the spin-orbit coupling, chemical potential and magnetic field) each nanowire may become a topological superconductor with Majorana zero energy modes located at its ends. Especially promising are hybrid systems consisting of epitaxial layers of a superconductor, a ferromagnetic insulator and a semiconductor, as reported in \cite{Marcus}. The experimental observations compatible with existence of  Majorana zero modes  in such systems are described, for example, in \cite{Marcus,Exp} and \cite{review}. 
In the present arrangement each MCB island contains four Majorana zero modes. A large charging energy $E_C$ fixes the charge and thereby encodes a qubit~\cite{Fu2010,BeriCooper2012}, where the two degenerate quantum states, $|\downarrow\ra$ ($|\uparrow\ra$), have $N_0$ ($N_0-2$) particles in the condensate and empty (filled) pairs of Majorana modes. 
 
  The idea to use arrays MCBs to produce exotic phases of matter has also been discussed in the literature \cite{Plugge,Pientka,Wille,Oreg,Oreg2,Z2}. However, in all these works the exchange interaction between the effective spins was the short range superexchange generated by the direct tunneling between the MCBs. In this paper I advocate for the a different arrangement where the dominant role is played by the Kondo screening and the Ruderman-Kittel-Kasuya-Yosida (RKKY) interaction.  
 
  The simplest arrangement fitting our purposes is to connect MCB to three external metallic leads via tunneling contacts leaving one Majorana zero mode idle. I depict such MCB as a triangle (see Fig. \ref{MCB}). The tunneling Hamiltonian is
 \bea
 H_{tun} = \exp(\ri\phi/2)\sum_{j}t_{j}\gamma_j\psi_j + \mbox{H.c.}, 
 \eea 
  where $\phi$ is the phase of the superconducting order parameter. Such tunneling process explicitly excludes the possibility of exciting quasiparticles \cite{Plugge2}. As was demonstrated in \cite{BeriCooper2012,Galpin2014}, if the charging energy of the box is much greater than the characteristic value of the tunneling matrix elements $E_C >> t_{j}$ one can integrate out the phase fluctuations of the condensate which results in the exchange interaction between such MCB and the electrons of the leads (it is supposed that they are spin polarized): 
 \bea
 H_{ex} =  J_K^{ij}(c^+_ic_j - c^+_jc_i)\gamma_i\gamma_j, ~~ J_K^{ij} \sim t_it_j/E_C, \label{Kondo}
 \eea
  where indices $i,j$ correspond to the leads. For a single MCB this interaction gives rise to the topological Kondo effect where the leads serve as the bulk.  The spin operator is 
  \be
  S^i = \frac{\rm i}{2}\epsilon_{ijk}\gamma_j\gamma_k, ~~ \{\gamma_k,\gamma_j\} = \delta_{jk}. \label{rep1}
  \ee
  There is an alternative representation taking in account that $\gamma_1\gamma_2\gamma_3\gamma_4 = 1/4$:
  \be
  S^i = \ri\gamma_4\gamma_i, ~~ i = 1,2,3. \label{rep2}
  \ee
   The physics of an array of MCBs is determined by two energy scales. One scale is the Kondo temperature $T_K \sim E_C\exp(- 1/J_K\rho)$, where $\rho$ is the electrons density of states, and the other is the characteristic RKKY exchange $J_{RKKY}(r) = \rho J_K^2 f(k_Fr)$, where $k_F$ is the Fermi wave vector and function $f(x)$ depends on the characteristics of the metallic wires connecting the MCBs. For purely one dimensional metallic wires $f(x) \sim \cos(2x)/|x|$, but in general it will depend on shape of the connecting wires.
  
\begin{figure}[!htb]
\centerline{\includegraphics[ angle = 0,
width=0.5\columnwidth]{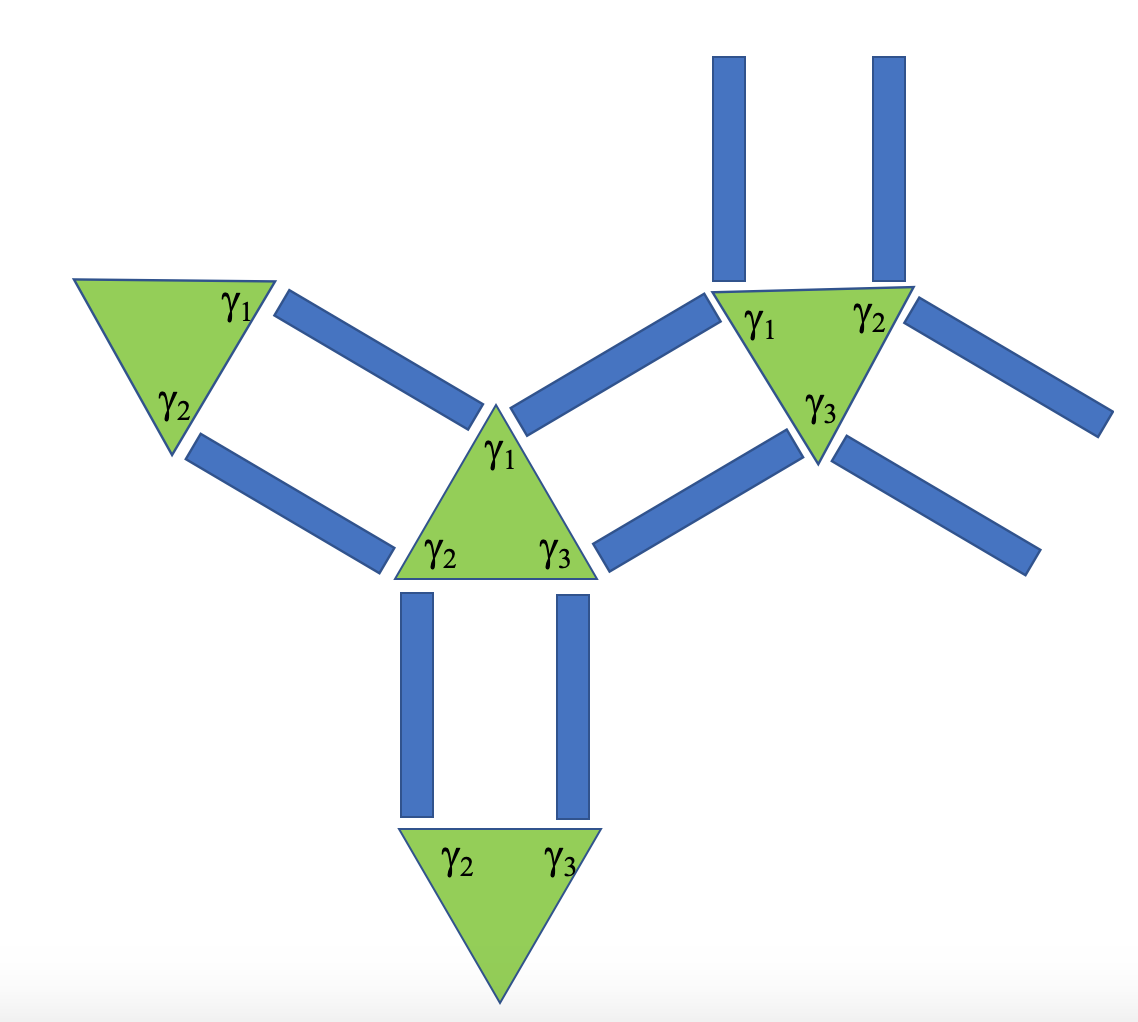}}
\vspace{-.50cm}
\caption{Lattice arrangement generating Kitaev interaction (\ref{kitaev}). The green triangles are MCBs containing effective spins 1/2, the blue lines are metallic wires. 
}
 \label{Kitaev}
\end{figure}

  {\it Kitaev model.} To obtain the Kitaev model we arrange MCBs on a hexagonal lattice connecting them by pairs of metallic leads as shown on Fig. \ref{Kitaev}. 
This model is essentially a Kondo lattice of effective spins made of Majorana zero modes. The pure spin model (\ref{kitaev}) is the limiting case  realized when the metallic wires connecting MCBs are short such that the RKKY interaction $J_{RKKY}$ dominates over the Kondo screening. Then the behavior of the system will be controlled by  the  RKKY exchange for bare spins. As I have mentioned above such situation has been described in the literature, most clearly in \cite{Pientka}. Due to the inherently bond-directed nature of the spin-fermion interaction (\ref{Kondo}) the RKKY exchange is also bond-directed as in (\ref{kitaev}).  So this  arrangement of MCBs and nanowires simulates the Kitaev model of spins S=1/2 where the excitations are propagating Majorana fermions and immobile gauge field fluxes. The difference is that in the original Kitaev model the physical observables are the spins and the Majorana fermions are nonlocal with respect to the spins being related to them via Jordan-Wigner transformation. In the present case one can follow the standard approach to the exact solution \cite{kitaev}  to establish that  these excitations are {\it local} with the respect to the MCB Majorana modes. Using (\ref{rep2}) we can see that paradoxically the propagating fermion is the "idle" $\gamma_4$ and the gauge field fluxes are made of contour products of other fermions. Indeed, substituting (\ref{rep2}) into (\ref{kitaev}) we obtain
\bea
&& H_{Kitaev} = \\
&&  \sum_{{\bf r},{\bf e}_{\mu}}J_{RKKY}({\bf r},{\bf e}_{\mu})\gamma_4({\bf r})\gamma_4({\bf r}+{\bf e}_{\mu})\Big[\gamma_{\mu}({\bf r})\gamma_{\mu}({\bf r} +{\bf e}_{\mu})\Big].\nonumber
\eea
The effective hopping matrix elements $t({\bf r}, {\bf r} + {\bf e}_{\mu}) = J_{RKKY}({\bf r},{\bf e}_{\mu})\Big[\gamma_{\mu}({\bf r})\gamma_{\mu}({\bf r} +{\bf e}_{\mu})\Big]$ are integrals of motion and hence the only propagating fermion is $\gamma_4$. Due to oscillatory character of the RKKY interaction one has to expect a fair amount of randomness in any practical realization of the suggested system. However, even in the presence of disorder the Kitaev model  remains  a model of free Majorana fermions. The recent numerical work on the model of fermions with random  velocity modulation \cite{Foster} suggests that the low energy states  are unaffected by the disorder, although some interesting multifractal physics develops at intermediate energies. 

{\it Kitaev Kondo lattice.} A different regime emerges when $T_K >> J_{RKKY}$ which is possible  when the wires are sufficiently long or thick.  Then the  Kondo effect will  develop and we will have a kind of a heavy fermion metal. This metal, however, will be  rather unusual  due to the non-Fermi liquid nature of the Kondo effect. As it was demonstrated in \cite{TopKondoWe}, the interaction (\ref{Kondo}) with three metallic  leads generates  the four-channel Kondo effect. The fermionic bilinear operator transforms as the $O_2(3) \equiv SU_4(2)$ Kac-Moody current, using  for the bulk the embedding  $O_1(6) = O_2 (3)\times O_3(2) = SU_4 (2)\times U(1)$ we arrive to the 4-channel Kondo effect with effective spin 1/2 coming from the MCB \cite{TopKondo,TopKondoWe}. The ground state of such Kondo impurity is quantum critical with correlation functions decaying in time with nontrivial power laws. It is important that the existence of the critical point is not affected by the anisotropy of the exchange interaction (\ref{Kondo}). Hence one should not worry about tunneling matrix elements at different leads to be unequal. At the critical point  the effective spins are not completely screened by the conduction electrons.  These critical remnants of the spins will interact with each other via the RKKY interaction and eventually some kind of new state will emerge at temperatures below $T_c\sim J_{RKKY}^3/T_K^2$,  there will be a region of non-FL quantum critical  behavior at $T_K >> T>>T_c$. Below I will try to determine some features of this state postponing a more detailed analysis to future publications. 
 
\begin{figure}[!htb]
\centerline{\includegraphics[ angle = 0,
width=0.3\columnwidth]{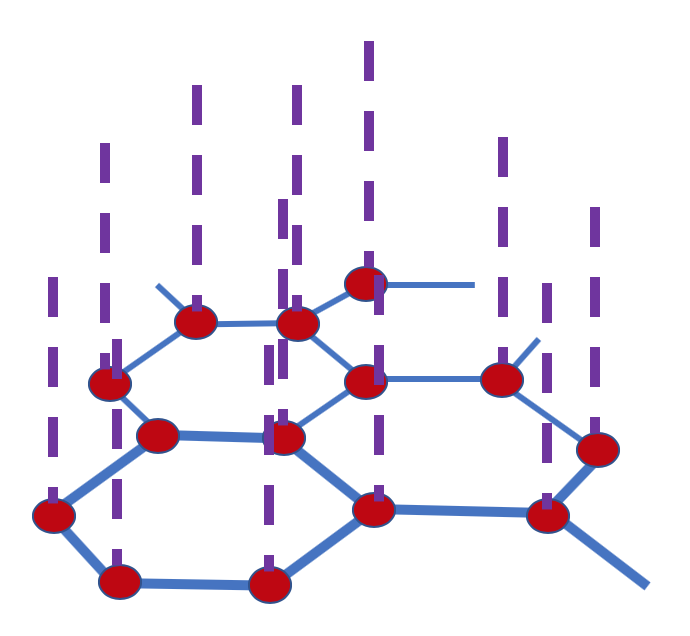}}
\vspace{-.50cm}
\caption{A cartoon depiction of the low energy description of the Kondo lattice described in the text. Red circles are overscreened  effective spins represented by vertex operators (\ref{vertex}) of a boundary conformal field theory. The fields $\phi_{1,2}$ of these theories propagate along  dashed purple lines.
}
 \label{monadas}
\end{figure}

  To enter into the regime described above one needs to reduce  the strength of the RKKY interaction. Since this interaction oscillates with distance as $\cos(2k_Fr)$, its reduction can be achieved  by varying the distance between the MCBs.  There are also other ways. In artificial systems like the one described here one can do this by increasing the volume of the conducting leads, which can be done if one uses, for example,  metallic disks instead of nanowires to connect MCBs.   If the condition $T_K >> J_{RKKY}$ is fulfilled then  at temperatures much smaller than $T_K$  the spin dynamics on each site can be described by the boundary conformal field theory  with central charge $C=2$ as is described in detail in \cite{TopKondoWe}.  This theory is  equivalent to the Gaussian model of  two chiral noninteracting bosonic fields:
 \bea
 S_0 = \sum_{a=1,2}\int \rd\tau \int_{-\infty}^\infty \rd y \p_y\phi_a(\p_y\phi_a + \ri \p_\tau\phi_a), \label{S0}
 \eea
 where $y$ is a fictitious coordinate.  To visualize this one can imagine that each MCB site is pierced by a line on which live two bosonic fields $\phi_{1,2}(y)$ as on  Fig.\ref{monadas} which dynamics is governed by the chiral Gaussian model. Then the spin fields are expressed as  \cite{TopKondoWe}: 
 \bea
 && S^z \sim \s^z\cos[\sqrt{8\pi/3}\phi_1(0)], \label{vertex}\\
 && S^{x,y} \sim \s^{x,y}\cos\Big\{\sqrt{2\pi/3}[\phi_1(0) \pm \sqrt 3\phi_2(0)]\Big\}, \nonumber
 \eea
where $\s^a$ are Klein factors- Pauli matrices and $0$ denotes the position of the field on the imaginary $y$-line. The observables (\ref{vertex}) have singular temporal correlations. The scaling dimension of these fields is 1/3 and the static susceptibility diverges as $\chi^{aa} \sim T^{-1/3}T_K^{-2/3}$.  So the resulting low energy model is similar to the Kitaev one, but the spin dynamics on each site is governed by action (\ref{S0}): 
\bea
S = \sum_{\bf r} \Big\{ S_0[\phi_{\bf r}]  + \sum_a J_{RKKY}^{(a)}({\bf r}) S^a({\bf r}) S^a({\bf r} + {\bf e}^a) \Big\}\label{action}
\eea
where ${\bf e}_{1,2} = (-1/2, \pm \sqrt 3/2), ~{\bf e}_3 = (1,0)$ and I assume that the interaction on different bonds can be different due to disorder and lattice irregularities. 

 To get a glimpse into the low energy state we may first consider the two-site problem which can be solved exactly. The bosonized form of the interaction $J_{RKKY}S^z_1S^z_2$  in this case is
\bea
&& V = J_{RKKY}(\s^z_1\s^z_2)\Big\{\cos[\sqrt{8\pi/3}(\phi_1 + \phi_2)] + \nonumber\\
&& \cos[\sqrt{8\pi/3}(\phi_1 - \phi_2)] \Big\}_0.
\eea
We can introduce new fields $\phi_{\pm} = (\phi_1 \pm \phi_2)/\sqrt 2$ and then the action splits into two independent integrable boundary sine-Gordon models with $\beta^2 = 16\pi/3$. The scaling dimension of the cosine term  is $2/3 <1$ meaning that the operator is relevant. The model enters strong coupling regime at $T_0 \sim |J_{RKKY}|^3/T_K^2$; at strong coupling the phases $\phi_{\pm}$ are pinned to the minimum of the cosine potential. To get the correlation functions one can just expand the cosine potential around the minimum such that the effective action becomes quadratic. Then a simple calculation gives  the following two point correlation functions  ($a= \pm$): 
\be
\la\la \phi_a(-\omega)\phi_a(\omega) \ra\ra \sim (|\omega| + \omega_0)^{-1}, ~~ \omega_0 \sim |J_{RKKY}|^3/T_K^2,  \label{2point}
\ee
The spin-spin correlation function changes the long time asymptotic so that at $\tau\omega_0 >> 1$ we have:
\be
\la\la S^a_i(\tau)S^a_j(0)\ra\ra \sim \frac{\delta_{ij}}{\tau^2}, ~~ a = x,y,z.
\ee
Please note, that the spin-spin correlation function remains local. This result persists when we consider a lattice of infinite number of sites. Perturbation theory expansion in $J_{RKKY}$ shows that the at low temperatures a singular behavior does not occur in the two-point functions; instead it  occurs in the four-point correlation functions of the spin operators \cite{SM}
 \bea
 &&\Gamma(\Omega,\omega,\omega';{\bf k}) = \sum_{\bf r} \re^{\ri {\bf k}({\bf r}- {\bf r}')}\times\\
 && \la\la S^a(\Omega +\omega,{\bf r})S^a(-\omega,{\bf r}) S^b(-\Omega +\omega',{\bf r}')S^b(\omega',{\bf r}')\ra\ra.\nonumber
 \eea
 Summation of the leading diagrams indicates that the singularity occurs simultaneously at all wave vectors and therefore is local in nature as in the two-site problem \cite{SM}. This feature indicates that the strong coupling physics will be robust with respect to disorder in $J_{RKKY}$. 
  I postpone a detailed discussion of this exotic low temperature regime till future publications. 

\begin{figure}[!htb]
\centerline{\includegraphics[ angle = 0,
width=0.5\columnwidth]{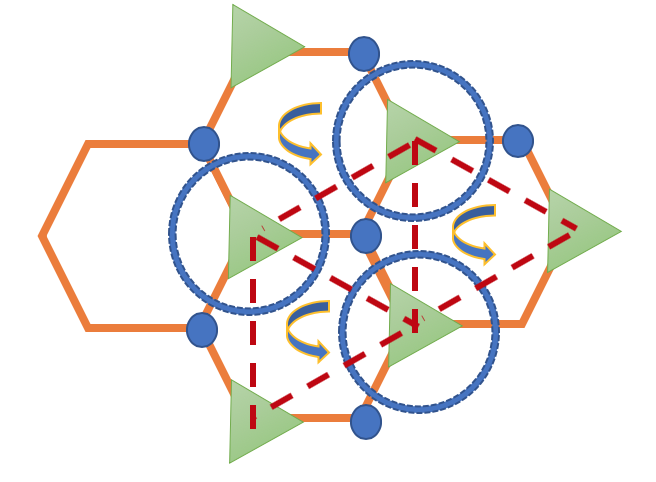}}
\vspace{-.50cm}
\caption{Each green triangle is a Majorana-Cooper box with three MZMs at the corners. It imitates spin 1/2. Each orange line is a metallic wire. Electrons from the wire can tunnel into the triangle giving rise to the exchange interaction (\ref{Kondo}).  The blue dots are tunneling matrix elements for electrons. The RKKY interaction (\ref{rkky}) takes place only among the effective spins placed at the corners of the lattice triangles marked by the circular arrow.
}
 \label{Spec}
\end{figure}

 {\it Three spin interaction.} To construct the three spin interaction (\ref{RKKY}) we will start with a model on  a hexagonal lattice where only half of the sites contain Majorana-Cooper pair boxes (Fig. \ref{Spec}). As in the case of Kitaev model, the links of the lattice are metallic wires. Electrons from a wire can tunnel into the MCB triangle giving rise to the exchange interaction (\ref{Kondo}). For sufficiently long wires this interaction gives rise to topological Kondo effect  which develops inside the area enclosed by the  blue dotted circle on Fig. \ref{Spec}. The blue dots are tunneling matrix elements connecting the areas where the Kondo effect takes place.  Apart from the Kondo effect there is RKKY exchange. It is easy to see that in the present arrangement it  (i) emerges only on the lattice triangles embedded into hexagons, and (ii) it necessarily includes three spins. However, the interaction is not SU(2) invariant as in Eq. (\ref{RKKY}), but  is given by 
  \bea
  H_{three}' = J_3 \sum_{{\bf r}} S^x({\bf r}+ {\bf e}_1)S^y ({\bf r}+ {\bf e}_2)S^z({\bf r}+ {\bf e}_3), \label{rkky}
 \eea
 where ${\bf r}$ is the center of the triangle made of the spins and ${\bf e}_i$ are three vectors pointing from the center to the corners. This is a highly frustrated interaction. Classically it would result in a highly degenerate ground state. The quantum case  requires careful investigation. 
 
\begin{figure}[!htb]
\centerline{\includegraphics[ angle = 0,
width=0.5\columnwidth]{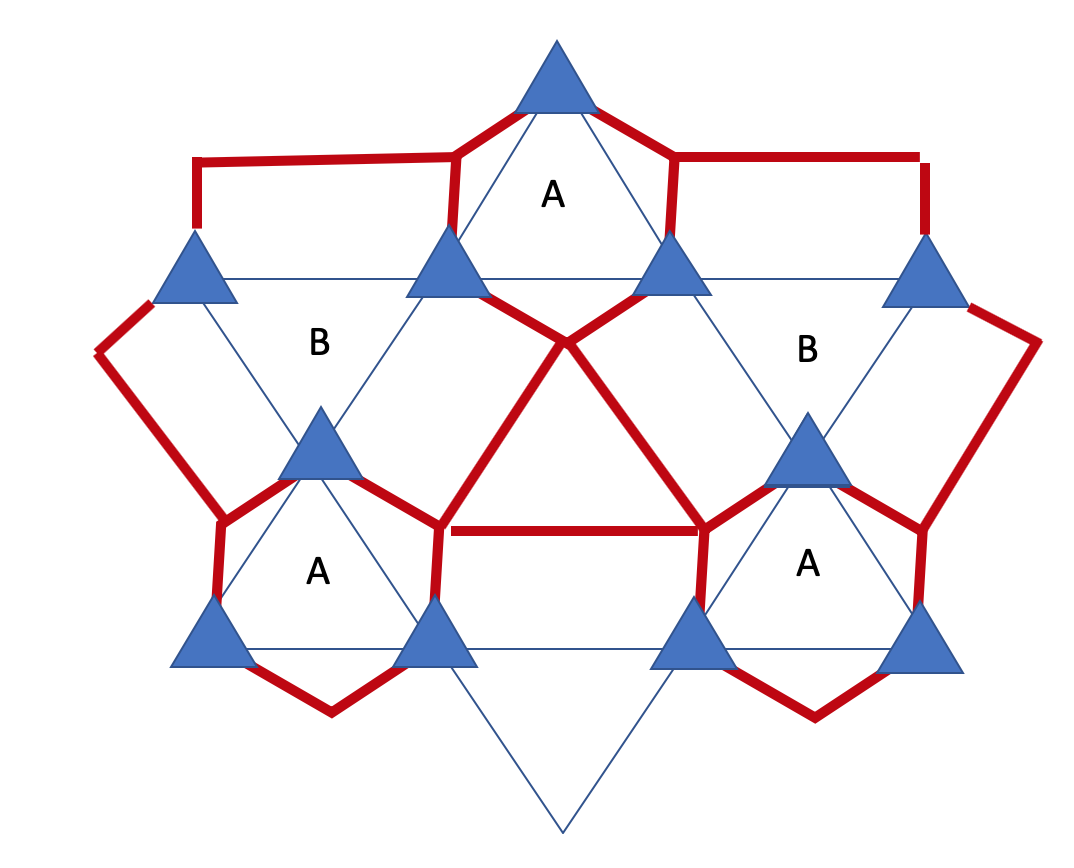}}
\vspace{-.50cm}
\caption{RKKY interaction on kagome lattice. The dark red lines are metallic wires.  
}
 \label{kagome}
\end{figure}

 More definite results can be obtained for a Kondo lattice model on  kagome lattice. The corresponding  arrangement  is shown on Fig. \ref{kagome}. There are two types of triangles here - A and B types and they require different wire patterns as is shown on the figure. The marked difference between this arrangement and the one on hexagonal lattice is that the same spin components take part in the interactions on A and B triangles. Therefore if one neglects the Kondo screening one gets essentially the classical model similar to Ising one, but with three spin interaction. In the limit of strong Kondo screening on each site we have decoupled charge and the spin sector with central charge $C=2$ which can be described by two Gaussian models. On each site one can chose one bosonic field to represent the spin component taking part in the interaction $S^z \sim \sin(\phi)$. 
 The other Gaussian field is decoupled and this remains critical. Then the action for these field components will be 
 \bea
 && S = S_0 + \\
 && J_3\sum_{\bf r}\sin[ \phi({\bf r}+ {\bf e}_1)]\sin[\phi({\bf r}+ {\bf e}_2)]\sin[\phi({\bf r}+ {\bf e}_3))].\nonumber
 \eea
 This  interaction is marginally relevant, its scaling dimension is 1. The ground state it will produce is some kind of spin liquid. The correlation functions of transverse components of the "spins" will be short ranged which can be established by perturbation theory expansion in $J_3$. 
 
{\it Conclusions and discussion. }  My suggestion is  that  periodic arrangements consisting of   Majorana-Cooper pair boxes (MCB) connected by  conducting bonds   can produce Kondo lattices with bond-directed interactions. The role of spins $S=1/2$ is played by two level systems made by Majorana zero modes located on MCBs. These Kondo lattices can exhibit various kinds of exotic physics. When  the RKKY interaction prevails over the Kondo screening these models effectively become spin ones with the famous  Kitaev model being one of them.  I find it quite remarkable that in the suggested  setting the excitations of the Kitaev model are local in terms of the Majorana fermions of the MCBs and hence can be directly assessed  in the experiments. On the other hand, when the RKKY exchange is relatively weak so that the Kondo effect is allowed to develop we may have all kinds of peculiar heavy fermion systems. This  springs from the fact that the Kondo effect in these systems is the four-channel one naturally leading to local quantum critical behavior on individual sites. Hence the  Kondo lattice with singular on-site temporal correlation functions fits perfectly into the Dynamical Mean Field theory (DMFT) paradigm when the physics is determined by local correlations. The results obtained indicate the existence of a non-Fermi liquid regime at $T_K > T > T_c \sim J^3_{RKKY}/T_K^2$ dominated by quantum critical fluctuations. Below $T_c$ the system enters into the strong coupling regime which detailed description will be given elsewhere.  

I am grateful to Laura Classen, Weiguo Yin and Robert Konik, Giniyat Khaliullin for a valuable remarks and to Reinhold Egger for a careful reading of the manuscript and valuable suggestions. This work was supported by U.S. Department of Energy (DOE) the Office of Basic Energy Sciences, Materials Sciences and Engineering Division under Contract No. DE-SC0012704.

\begin{widetext}
\section{Supplementary Material}

\subsection{Correlation functions}
Here we sum the leading diagrams for the irreducible four point correlation function
\bea
\Gamma^{ab}(1,2,3,4;{\bf k}) = \sum_{{\bf r}} \re^{\ri {\bf k r}}\la\la S^a(\tau_1,{\bf r})S^a(\tau_2,{\bf r}) S^b(\tau_3,0)S^b(\tau_4,0)\ra\ra
\eea
The bare correlation functions are all local in space and equal to ($\beta^2 = 8\pi/3)$: 
\bea
&& \gamma_{zz}(1,2,3,4) = \la\la \cos\beta\phi_1(1)\cos\beta\phi_2(2)\cos\beta\phi_1(3)\cos\beta\phi_1(4)\ra\ra \sim \nonumber\\
&& =  s(\tau_{12})^{-2/3}s(\tau_{34})^{-2/3}\Big\{\Big[\frac{s(\tau_{13})s(\tau_{24})}{s(\tau_{14})s(\tau_{23})}\Big]^{1/3} -  \Big[\frac{s(\tau_{14})s(\tau_{23})}{s(\tau_{13})s(\tau_{24})}\Big]^{1/3}\Big\}^2 + \nonumber\\
&& s(\tau_{13})^{-2/3}s(\tau_{24})^{-2/3}\Big\{\Big[\frac{s(\tau_{12})s(\tau_{34})}{s(\tau_{14})s(\tau_{23})}\Big]^{2/3} -1\Big\} - s(\tau_{14})^{-2/3}s(\tau_{23})^{-2/3}. 
\eea
$\gamma_{zz} = \gamma_{xx} = \gamma_{yy}$, and 
\be
s(\tau) = \frac{\sin(\pi T\tau)}{\pi T}.\nonumber
\ee

We will parameterize the times as follows:
\bea
\gamma_{ab}(\tau/2, -\tau/2; \tau_0+\tau'/2, \tau_0-\tau'/2) 
\eea

Then in the limit of zero temperature we have 
\bea
&& \gamma_{zz}^{(1)} = \frac{1}{\tau^{2/3}\tau'^{2/3}}\Big\{\Big|\frac{\tau_0^2-\tau_-^2}{\tau_0^2-\tau_+^2}\Big|^{1/3} - \Big|\frac{\tau_0^2-\tau_+^2}{\tau_0^2-\tau_-^2}\Big|^{1/3}\Big\}^2, \label{zz1}\\
&&\gamma^{(2)}_{zz} = \frac{(\tau_+^2 - \tau_-^2)^{2/3}}{(\tau_0^2 -\tau_+^2)^{2/3}(\tau_0^2 - \tau_-^2)^{2/3}} - \frac{1}{(\tau_0^2 - \tau_+^2)^{2/3}}- \frac{1}{(\tau_0^2 - \tau_-^2)^{2/3}}\label{zz2}.
\eea

\bea
&& \gamma_{zx}(1,2,3,4) = \la\la \s^z(1)\cos\{\beta\phi_1(1)\}\s^z(2)\cos\{\beta\phi_2(2)\}\s_x(3)\cos\{\beta[\phi_1(3)+\sqrt 3\phi_2(3)]/2\}\s_x(4)\cos\{\beta[\phi_1(4)+\sqrt 3\phi_2(4)]/2\}\ra\ra \sim \nonumber\\
&& s(\tau_{12})^{-2/3}s(\tau_{34})^{-2/3}\Big\{\Big|\frac{s(\tau_{13})s(\tau_{24})}{s(\tau_{14})s(\tau_{23})}\Big|^{1/6} -\Big|\frac{s(\tau_{14})s(\tau_{23})}{s(\tau_{13})s(\tau_{24})}\Big|^{1/6}\Big\}^2  \rightarrow \nonumber\\
&& \frac{1}{\tau^{2/3}\tau'^{2/3}}\Big\{\Big|\frac{\tau_0^2-\tau_-^2}{\tau_0^2-\tau_+^2}\Big|^{1/6} - \Big|\frac{\tau_0^2-\tau_+^2}{\tau_0^2-\tau_-^2}\Big|^{1/6}\Big\}^2. \label{xz}
\eea
$\gamma_{zx} = \gamma_{z,y} = \gamma_{xy}$.

\subsection{Calculations of Fourier transforms}

We will consider the zero temperature limit first.  We will also consider the limit of zero incoming frequency $\Omega =0$. Let us introduce new variables $a = |\tau_+/\tau_-|^{1/2},  ~~\rho = |\tau_+\tau_-|^{1/2}$. 
Then at small $a<<1$  integral over $\tau_0$ in (\ref{zz1}) yields :
\bea
\int \rd \tau_0 \gamma_{xz} (\tau_0) \approx 
 \rho^{-1/3}\Big[\frac{\Gamma(5/6)\Gamma(5/6)}{\Gamma(5/3)}a + 1.86 a^{1/3} +...\Big].  \label{xz} 
\eea
To do the rest of integration we replace   $\chi = \log a$ and, taking into account  that the integrand (\ref{xz})  vanishes at $\chi =0$, we can approximate the Fourier transform of $\gamma_{xz}$ as follows: 
\bea
&& F_{xz}(\omega,\omega') \approx (4/3) \int \rho \rd \rho \rho \rd\chi \frac{\cos[\rho(\omega_+\re^{\chi} + \omega_-\re^{-\chi})]}{ \rho^{1/3}} \frac{(\re^{|\chi|} -1)}{[4\sinh(2\chi)]^{2/3}},
\eea
Let $\omega_+ = \Omega\re^{\phi}, ~~\omega_- = \Omega\re^{-\phi}$. Then (see Fig. \ref{fphi}): 
\bea
&& F_{xz}(\Omega,\phi) =  A_{xz} |\Omega|^{-5/3}f(\phi), \\
&&  f(\phi) \approx 1.5 \re^{-|\phi|/3}, ~~ |\phi| >>1, \label{Fxz} 
\eea
where $f(|\phi|)$ is depicted on Fig. \ref{fphi}. 

\begin{figure}[!htb]
\centerline{\includegraphics[ angle = 0,
width=0.5\columnwidth]{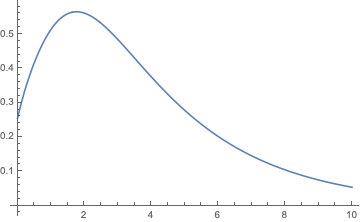}}
\vspace{-.50cm}
\caption{$f(|\phi|)$
}
 \label{fphi}
\end{figure}
\bea
 \int \rd \tau_0 \gamma^{(1)}_{zz} (\tau_0) \approx 
 \rho^{-1/3}\Big[\frac{\Gamma(1/2)\Gamma(1/3)}{\Gamma(5/6)}a^{-1/3} + 12.65 a^{1/3} +...\Big],\label{zz1}\\
\eea
Integral over $\tau_0$ in (\ref{zz2}) yields 
\bea
&&\frac{1}{\rho^{1/3}}{\cal F}(a), ~~ {\cal F}(a) = 2\Big[\int_0^{\infty} \frac{\rd x (a^2-1/a^2)^{2/3}}{(x^2 -a^2)^{2/3}(x^2 - a^{-2})^{2/3}} - (a^{1/3}+ a^{-1/3})\int_0^{\infty}\frac{\rd x}{(x^2-1)^{2/3}} \Big].
\eea
We have ${\cal F}(a) = {\cal F}(1/a)$/ For $a <1$ we have 
\bea
&& {\cal F}(a) = (a^{-2} -a^2)^{2/3}\Gamma(1/3)\Big\{\frac{\Gamma(1/2)}{\Gamma(5/6)}a F(2/3,1/2,5/6;a^4) + \frac{\Gamma(5/6)}{\Gamma(7/6)}a^{5/3}F(2/3,5/6,7/6;a^4) + \nonumber\\
&& a(1-a^4)^{-1/3}\Big[ \frac{\Gamma(-1/6)}{\Gamma(1/6)}a^{2/3}F(1/2,1/3,7/6;a^4) + \frac{\Gamma(1/6)}{\Gamma(1/2)}F(1/3,1/6,5/6;a^4)\Big]\Big\} - \nonumber\\
&& 3(a^{1/3} + a^{-1/3})\frac{\Gamma(1/2)\Gamma(1/3)}{\Gamma(5/6)} \\
&& \approx - \frac{\Gamma(1/2)\Gamma(1/3)}{\Gamma(5/6)}a^{-1/3}(2 +a).
\label{zz2}
\eea

Let us add up (\ref{zz1},\ref{zz2}) and denote $\exp\chi = a$, then integral over $\rho = (\tau_+\tau_-)^{1/2}$ yields 
\bea
&& \int \frac{\rd\chi }{|\omega_+\re^{\chi} + \omega_-\re^{-\chi}|^{5/3}}{\cal F}(\chi) = (\omega_+\omega_-)^{-5/6}F(|\phi|), ~~ \re^{\phi} = (\omega_+/\omega_-)^{1/2},\\
&& F(|\phi|) \approx - \re^{|\phi|/3}\Big(\frac{\Gamma(1/2)\Gamma(1/3)}{\Gamma(5/6)} + 12.5 \re^{-|\phi|} +...\Big) \label{ZZ}
\eea

Summing up the results (\ref{Fxz},\ref{ZZ}) we get the following asymptotic: 
\bea
&& \gamma_{zz}(\omega_+,\omega_-) \sim - \Big[\mbox{max}|\omega_a|\Big]^{-2/3}\Big[\mbox{min}|\omega_a|\Big]^{-1}, \nonumber\\
&& \gamma_{xz}(\omega_+,\omega_-) \sim + \Big[\mbox{min}|\omega_a|\Big]^{-2/3}\Big[\mbox{max}|\omega_a|\Big]^{-1}.\label{factor}
\eea

\subsection{The propagator}
\begin{figure}[!htb]
\centerline{\includegraphics[ angle = 0,
width=0.5\columnwidth]{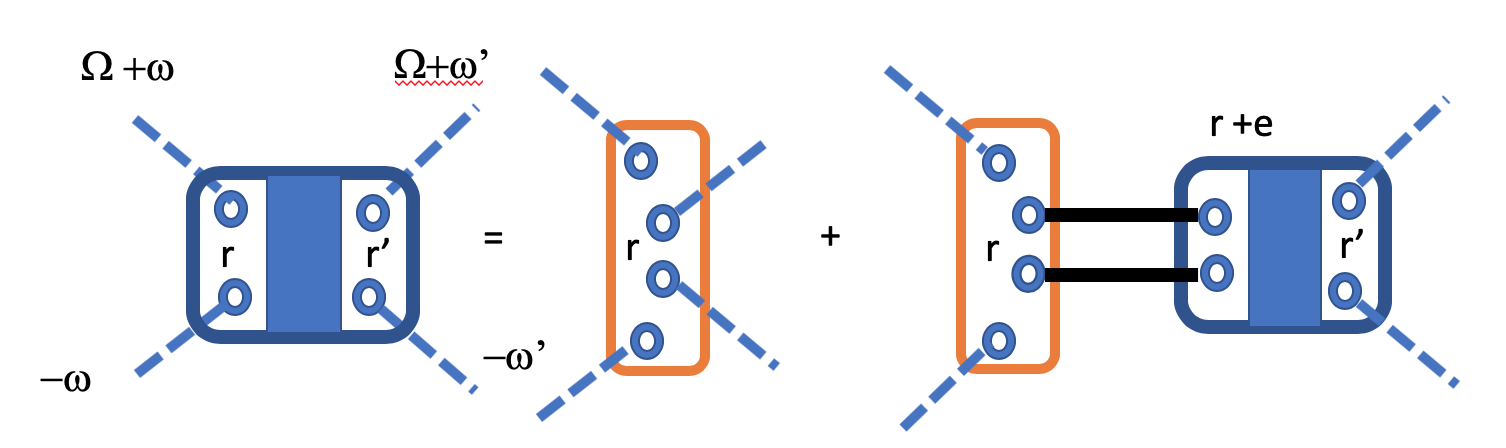}}
\vspace{-.50cm}
\caption{Summation of the ladder diagrams. The blue rectangles are $\Gamma$'s, the orange ones are the bare irreducible four-point functions $\gamma$, the circles are spin operators, the black solid lines are the exchange integrals.  
}
 \label{BS}
\end{figure}

The summation of the leading diagrams in the perturbation series give the following equation (Fig. \ref{BS}): 
\bea
\Gamma_{ab}^{AB}(1,2;3,4,{\bf k}) = \delta_{AB}\gamma_{ab}(1,2;3,4) + J^2\int \rd \eta_1\rd\eta_2 \gamma_{ac}(1,2;\eta_1,\eta_2)\Gamma^{\bar A,B}_{cb}(\eta_1,\eta_2;3,4,{\bf k})\re^{\ri{\bf k}{\bf e}_c}, \label{Gamma}
\eea
where $A,B$ stand for sublattice and $\bar A$ means sublattice different from $A$.

This equation can be solved approximately by Fourier transformation taking into account that the vertex can be approximately factorized as in (\ref{factor}). Then assuming that $F$ is a slow function of the frequencies one can approximate the integral:  
\bea
&& \int \rd \omega' F(\omega',\omega_2)\gamma_{zz}(\omega,\omega') \approx \nonumber\\
&& \int _0^{\infty}\rd \omega'\frac{F(\omega'; \omega_2)}{(\omega+\omega')^{2/3}|\omega -\omega'|} + \int_{-\infty}^0 \rd\omega' \frac{F(\omega'; \omega_2)}{|\omega+\omega'|^{2/3}(\omega -\omega')}\nonumber\\
&&\frac{1}{\omega^{2/3}} \int \rd x \frac{F(\omega x) + F(-\omega x)}{(1+x)^{2/3}|1-x|} \sim -4\frac{\ln(\omega/T)}{ 2\omega^{2/3} }F(\omega). 
\eea
A similar calculation can be performed for $\gamma_{xz}$ which yields  a similar answer, but without the logarithm.

 Then the integral equation (\ref{Gamma}) becomes algebraic:
\bea
\left(
\begin{array}{cccccc}
1 & 0 & 0 &g\re^{\ri k_1} & \bar g \re^{\ri k_2} & \bar g \re^{\ri k_3}\\
0 & 1 & 0 & \bar g\re^{\ri k_1} & g \re^{\ri k_2} & \bar g \re^{\ri k_3}\\
0 & 0 & 1 & \bar g\re^{\ri k_1} & \bar g \re^{\ri k_2} & g \re^{\ri k_3}\\
g\re^{-\ri k_1} & \bar g \re^{-\ri k_2} & \bar g \re^{-\ri k_3} & 1 & 0 &0\\
\bar g\re^{-\ri k_1} &  g \re^{-\ri k_2} & \bar g \re^{-\ri k_3} & 0 & 1 &0\\
\bar g\re^{-\ri k_1} &  \bar g \re^{-\ri k_2} &  g \re^{-\ri k_3} & 0 & 0 &1
\end{array}
\right)\left(
\begin{array}{c}
\Gamma_1^A\\ \Gamma_2^A\\ \Gamma_3^A \\ \Gamma_1^{\bar A} \\ \Gamma_2^{\bar A}\\ \Gamma_3^{\bar A}
\end{array}
\right) = \hat g.
\eea
where 
\bea
g = - g_0|\omega|^{-2/3}\log(|\omega|/T), ~~ \bar g = g_1|\omega|^{-2/3},
\eea
with $g_{0,1} \sim J^2_{RKKY}$. The determinant of this matrix is 
\bea
[1 - (g-\bar g)^2]\Big\{1 - [1-(g-\bar g)^2](g + 2\bar g)^2 + 4\bar g^2\Big[\sin^2 k_x/2 + \sin^2(3k_x/4 + \sqrt 3 k_y/4) +  \sin^2(3k_x/4 - \sqrt 3 k_y/4)\Big]\Big\}.
\eea
As we can see, the momentum dependent factor remains positive when the  first bracket vanishes. This is an indication that the strong coupling limit does not develop long range correlations. The frequency at which the vertex goes to strong coupling is 
\bea
\omega^{2/3} \sim J^2_{RKKY}.
\eea

\end{widetext}

\end{document}